\documentclass[10pt,letterpaper]{article}
\usepackage{opex3}
\usepackage{bm,amsmath,amssymb}
\usepackage{cite}
\usepackage[caption=false]{subfig}

\newcommand{\phiE}{\phi(-\rho| P_{Z|X,H_E},P_X)}
\newcommand{\Hc}{H_{\mathrm{sec}}(R_E)}

\begin{document}
\bibliographystyle{osajnl}
\title{Free-space optical channel estimation for physical layer security}

\author{
Hiroyuki~Endo,$^{1,2}$ 
Mikio~Fujiwara,$^1$ 
Mitsuo~Kitamura,$^1$ 
Toshiyuki~Ito,$^1$
Morio~Toyoshima,$^3$
Yoshihisa~Takayama,$^{3,4}$
Hideki~Takenaka,$^3$
Ryosuke~Shimizu,$^5$
Nicola~Laurenti,$^6$
Giuseppe~Vallone,$^6$
Paolo~Villoresi,$^6$
Takao~Aoki$^2$ and Masahide~Sasaki$^{1,*}$}%

\address{$^1$ Quantum ICT Laboratory, National Institute of Information and Communications Technology, Koganei, 184-8795, Japan\\
         $^2$ Department of Applied Physics, Waseda University, Shinjuku, 169-8050, Japan\\
         $^3$ Space Communication Systems Laboratory, National Institute of Information and Communications Technology, Koganei, 184-8795, Japan\\
         $^4$ Current affiliation: School of information and Telecommunication Engineering, Tokai University, Takanawa, Minato, 108-8619, Japan\\
         $^5$ Center for Frontier Science and Engineering, the University of Electro-Communications, Chofu, 182-8585, Japan\\
         $^6$ Department of Information Engineering, University of Padova, via Gradenigo 6/B, 35131 Padova, Italy}

\email{$^*$psasaki@nict.go.jp}

\begin{abstract}
  We present experimental data on message transmission in a free-space optical (FSO) link at an eye-safe wavelength, 
  using a testbed consisting of one sender and two receiver terminals, where the latter two are a legitimate receiver and an eavesdropper.  
  The testbed allows us to emulate a typical scenario of physical-layer (PHY) security such as satellite-to-ground laser communications.  
  We estimate information-theoretic metrics including secrecy rate, secrecy outage probability, and expected code lengths for given secrecy criteria based on observed channel statistics.  
  We then discuss operation principles of secure message transmission under realistic fading conditions, and provide a guideline on a multi-layer security architecture by combining PHY security and upper-layer (algorithmic) security.  
\end{abstract}


\section{Introduction}
  Free-space optical (FSO) communication is a promising technology for enhancing the connectivity of wireless networks \cite{chanFSO}, 
  thanks to the features such as wide band width in an unregulated spectrum, ultra-low inter-channel interference, and power-efficient transmission.  
  Potential applications of FSO communication include satellite laser communications \cite{Toyoshima}, 
  a network system comprised of unmanned aerial vehicles, high-altitude platforms or drones \cite{UAVFSO, HAPFSO, facebook}, 
  the last one mile link from the fiber backbone to the clients premises \cite{urban}  
  and military applications \cite{Military, navalFSO}.   
  
  As FSO communication becomes more and more important in these applications, the security requirements also become more demanding. 
  Although high directionality of laser beam makes FSO communication inherently more secure than RF counterparts, 
  FSO communication can still suffer from optical tapping risks, especially when the main lobe of laser beam footprint is considerably wider than the receiver size \cite{Agaskar, Puryear, Eghbaldeff}.   
  These risks would be pronounced in urban communications where an eavesdropper would hide in the top of the same building as the legitimate receiver \cite{LopezMartinez},
  or in satellite-to-ground laser communications in which the beam footprint would be a scale of km.  
  Therefore, secure communication over FSO links still remains a challenging task.  
  
  Traditionally, security has been highly dependent on the upper layer protocols 
  such as conventional encryption techniques with a pre-shared secret key or a key exchanged via public key cryptosystems.  
  The security of these protocols is proved with algorithmic means.  
  Then, it will be weakened as computer technologies and decryption algorithms are advancing.  
  Moreover, with the rapid growth of the number of communication nodes, the key distribution and management are becoming increasingly difficult, and are introducing larger overhead and latency to the system. 
  During the past few years, however, physical layer (PHY) security \cite{BBBook, Zhoubook} has been gaining research attentions as a means to complement conventional encryption techniques.  
  Its security is provided in information theoretical manners based on the particular coding techniques \cite{wiretapreview, WynerWiretap, csiskor, AhlswedeSKA, MaurerSKA} 
  or the careful signal designs \cite{MIMOMEreview, negiartificialnoise, KhistiMIMOSE, OggierMIMO}.  
  Different from conventional encryption techniques, no computational assumptions are placed on the eavesdropper. 
  Practically, the existing security system can be enhanced by introducing PHY security as a first line of defense against eavesdropping.  
  
  The fundamental theoretical frameworks of PHY security was laid by Wyner \cite{WynerWiretap}, and Csisz\'ar and K\"orner \cite{csiskor} based on secure message transmission over a wiretap channel, 
  and by Ahlswede and Csisz\'ar \cite{AhlswedeSKA}, and Maurer \cite{MaurerSKA} based on secret key agreement from common randomness.  
  In the wiretap channel model, its security is provided by an appropriate channel code guaranteeing both the reliability for the legitimate receiver and the secrecy against the eavesdropper.
  If the channel from the sender to the eavesdropper is a degraded version of that from the sender to the legitimate receiver, 
  a non-zero secrecy rate can be achieved by sacrificing a fraction of the message rate.  
  While such a degraded condition seems not to be realistic in wired communications, 
  it is more reasonable in FSO communications built up between parties with a direct line-of-sight (LoS).  
  Specifically, in LoS links, surveillance cameras will be able to detect any suspicious activity which makes it harder for an eavesdropper to intercept the main lobe of laser beam. 
  On the perspective of this scenario, there have been theoretical studies \cite{LopezMartinez, Wang, numsec, oamfsophysec} on the potential of PHY security in FSO communications.  
  
  In spite of these remarkable theoretical studies, realization of PHY security is still a challenging issue. 
  In FSO links, the intensity of the transmitted beam and the statistics of the received signals vary in a time scale of ms due to atmospheric turbulences.  
  This kind of fading effect makes it difficult to implement an efficient coding scheme which can ensure PHY security under various conditions.  
  Some approaches to mitigate the effect of atmospheric turbulences have been investigated such as the adaptive real time selection technique \cite{paoloturb2} 
  in a horizontal quantum communication link of 143 km between La Palma and Tenerife Islands \cite{paoloturb}.  
  However, experimental data and analyses on FSO fading links from the viewpoint of PHY security are still overwhelmingly lacking.  
  
  This motivates us to collect transmission data in an FSO wiretap channel and analyze them in terms of fundamental metrics of PHY security.  
  To this end, we have constructed a metropolitan terrestrial FSO link testbed (Tokyo FSO Testbed).
  This testbed consists of one sender terminal and two receiver terminals, one for the legitimate receiver and the other for the eavesdropper.  
  Each of the two is 7.8 km distance apart from the sender terminal.  
  The purposes of the testbed are 
  (1) to examine PHY security techniques (e.g., secure message transmission and secret key agreement) in real-field FSO links, 
  (2) to emulate typical FSO communication scenarios such as satellite-to-ground laser communications, 
  and (3) to accumulate transmission data under several real-field conditions and utilize them for practical system design.  
  In this paper, we focus on secure message transmission and analyze the characteristics of the FSO wiretap channel by transmitting a pseudorandom binary sequence based on on-off keying modulation. 
  Using the output signal statistics, we estimate secrecy rates and related security metrics. 
  We then discuss how the legitimate party can set a guideline for operating secure message transmission based on the observed data with pilot signals and the data accumulated from the past. 
  
\section{Wiretap channel and performance metric} \label{sec3}
      \begin{figure}[tb]
        \centering
        \includegraphics[width=7cm, bb=0 0 994 704]{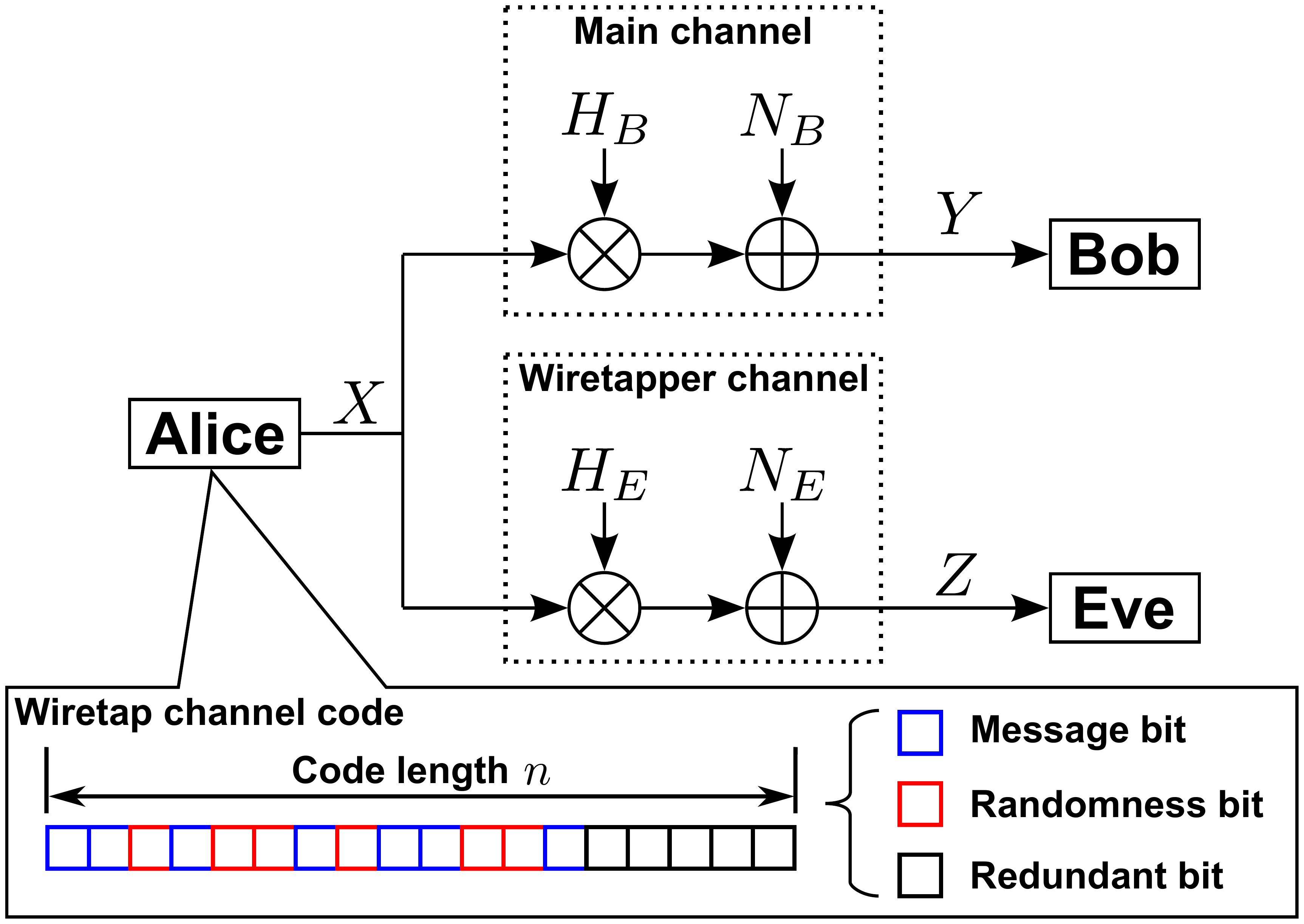}
        \caption{Schematic diagram of the wireless wiretap channel model and wiretap channel coding \cite{SAT}. } \label{wiretapchannelmodel}
      \end{figure}
    Throughout the paper, we consider secure message transmission via a wireless wiretap channel system illustrated in Fig. \ref{wiretapchannelmodel}.  
    The sender (Alice) encodes a confidential message into a code word random variable (RV) $X^n$ for transmission over the wiretap channel, where $n$ is the code length.  
    We assume that Alice uses on-off keying modulation, thus RV $X$ takes a value with $0$ or the peak power of the transmission laser.  
    Moreover, we also assume that the laser power and the input probability distribution $P_X$ over $X$ are fixed irrespective of the channel state.  
    
    The legitimate receiver (Bob) observes the output via a discrete-time quasi-static fading channel (the main channel) given by
      \begin{align}
        Y = H_B X+ N_B,
      \end{align}
    being $H_B$ the channel gain RV and $N_B$ the additive white Gaussian noise (AWGN) RV.  
    The eavesdropper (Eve) is also capable to observe Alice's transmission from the output via a discrete-time quasi-static fading channel (the wiretapper channel) given by
      \begin{align}
        Z = H_E X+ N_E, 
      \end{align}
    being $H_E$ the channel gain RV and $N_E$ the AWGN RV.  
    Here, upper case letters $X,Y,Z, H_B, H_E$ are all positive real RVs since we modulate intensity of light.  
    For later convenience, we shall use lower case letters $x,y,z,h_B,h_E$ to denote realizations of $X,Y,Z,H_B,H_E$, respectively.  

    In order to transmit $m$ bits of confidential information reliably and securely through the wiretap channel, 
    Alice introduces some redundancy to perform error correction and some random dummy information as the cost of additional secrecy.  
    This entails adding redundant bits and $l$ random dummy bits to the code word and hence increasing its length to $n$ as shown in the lower panel of Fig. \ref{wiretapchannelmodel}.
    This scheme is particularly referred to as wiretap channel coding.   
    To design a wiretap channel code of length $n$, we shall specify two rates, the message rate $R_B = m/n$ and the randomness rate $R_E = l/n$ in bits/letter.  

    We assume that the channel gains $H_B$ and $H_E$ remain constants $h_B$ and $h_E$, respectively, during specific interval (coherence interval).
    In each coherence interval with channel realizations $h_B$ and $h_E$, 
    one has to satisfy $R_B + R_E \le I(P_X,P_{Y|X,H_B})$ to establish a reliable communication, 
    where $I(P_X,W)$ is the mutual information with the input probability distribution $P_X$ and the transition probability distribution $W$, 
    and $P_{Y|X,H_B}$ denotes the transition probability distribution of the main channel.  
    On the other hand, $R_E \ge I(P_X,P_{Z|X,H_E})$ should be satisfied for confidentiality, 
    where $P_{Z|X,H_E}$ denotes the transition probability distribution of the wiretapper channel.  
    Hence, the message rate $R_B$ must satisfy the following relation in each coherence interval,
      \begin{align}
        R_B \le R_\mathrm{S,i}(h_B, h_E) \equiv \max[0,I(P_X,P_{Y|X,H_B}) - I(P_X,P_{Z|X,H_E})].   \label{secrate}
      \end{align}
    We shall call $R_\mathrm{S,i}(h_B, h_E)$ the instantaneous secrecy rate given $h_B$ and $h_E$.  
    When there is no ambiguity, we will drop the dependence on $(h_B, h_E)$.  
    
    In this paper, Bob is assumed to use hard-decision decoding where the value of individual bits are quantized to either $y = 0$ or $y = 1$ based on the threshold $y_\mathrm{th}$.  
    Thus, the mutual information $I(P_X,P_{Y|X,H_B})$ is calculated as
      \begin{align}
        I(P_X,P_{Y|X,H_B}) = \sum_{x \in \{0,1\}} \sum_{y\in \{0,1\}} P_X(x) P_{Y|X,H_B}(y|x,h_B) \log_2 \left[\frac{P_{Y|X,H_B}(y|x,h_B)}{\sum_{x^{\prime}}P_X(x^{\prime}) P_{Y|X,H_B}(y|x^{\prime},h_B)} \right].  
      \end{align}
    In the above equation, the transition probability functions for $y=1$ and $y=0$ given by $x \in \{0,1\}$ are defined as
      \begin{align}
        P_{Y|X,H_B}(1|x,h_B) = \frac{N(y \ge y_\mathrm{th}|x,h_B)}{N(x)}, \quad P_{Y|X,H_B}(0|x,h_B) = \frac{N(y \le y_\mathrm{th}|x,h_B)}{N(x)}, \label{apostehard}
      \end{align}
    where $N(x)$ is the number of an input $x \in \{0,1\}$ transmitted by Alice, 
    and $N(y \ge y_\mathrm{th}|x,h_B)$ and $N(y \le y_\mathrm{th}|x,h_B)$ 
    are the numbers of events of $y \ge y_\mathrm{th}$ and $y \le y_\mathrm{th}$ conditioned by an input $x\in \{0,1\}$ in the coherence interval with the channel realization $h_B$, respectively. 
    The threshold $y_\mathrm{th}$ should be numerically optimized such that the mutual information $I(P_X,P_{Y|X,H_B})$ is maximized.  
    
    On the other hand, Eve is assumed to use soft-decision decoding which uses a whole range of output values to make decisions.  
    This is the reasonable in secure communication, since the mutual information based on soft-decision decoding is slightly larger than that based on hard-decision decoding \cite{proakis}.  
    In this decoding, considering the finite size of samples (see Appendix A), 
    we quantize the experimental data into $K$ bins with an identical width $\Delta$.  
    Thus, the transition probability function $P_{Z|X,H_E}(z^{(i)}|x,h_E)$ is calculated as follows:
      \begin{align}
        P_{Z|X,H_E}(z^{(i)}|x,h_E) = \frac{N(z^{(i)}|x,h_E)}{N(x)}, \label{apostesoft}
      \end{align}
    being $N(z^{(i)}|x,h_E)$ the number of events that $z$ is in $i$-th bin conditioned by an input $x \in \{0,1\}$ in the coherence interval with the channel realization $h_E$.  
    The mutual information $I(P_X,P_{Z|X,H_E})$ is calculated as
      \begin{align}
        I(P_X,P_{Z|X,H_E}) = \sum_{x \in \{0,1\}} \sum^K_{i=1} P_X(x) P_{Z|X,H_E}(z^{(i)}|x,h_E) \log_2 \left[\frac{P_{Z|X,H_E}(z^{(i)}|x,h_E)}{\sum_{x^{\prime}}P_X(x^{\prime}) P_{Z|X,H_E}(z^{(i)}|x^{\prime},h_E)} \right].  \label{muteve}
      \end{align}
    
\section{Overview of the experimental setup} \label{sec2}
  The ideal technological goal is to evaluate the instantaneous secrecy rate $R_\mathrm{S,i}$ 
  by monitoring the channel state information (CSI), namely, the channel gains of both the main and wiretapper channels, 
  in various conditions depending on weather, temperature, and instruments.  
  However, due to the effect of atmospheric turbulences, the gains $H_B$ and $H_E$ are always fluctuating and hardly predicted.  
  This kind of fading effect makes it difficult to implement an efficient wiretap channel coding scheme under various conditions.  
  This motivates us to collect transmission data in an FSO wiretap channel by using Tokyo FSO Testbed introduced in this section.  
  The testbed and its experimental data allow us not only to analyze the secrecy performance of secure message transmission based on the information-theoretic metrics, 
  but also to determine whether secure message transmission can be established or not under a given condition.  
  
    \begin{figure}[tb]
        \centering
          \subfloat{
            \includegraphics[width=4cm, bb=0 0 580 802]{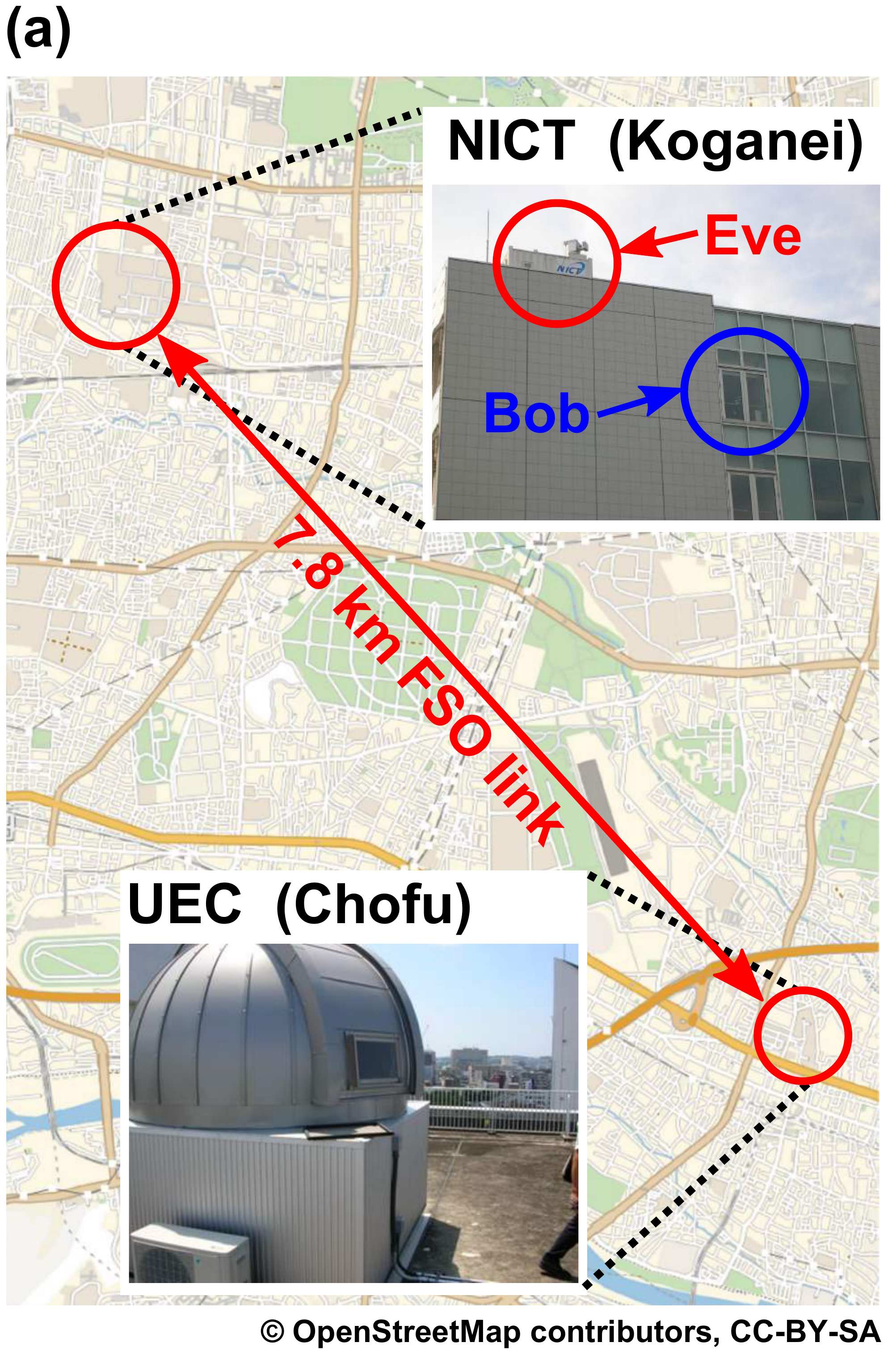}
          }
          \subfloat{
            \includegraphics[width=8.5cm, bb=0 0 1672 1138]{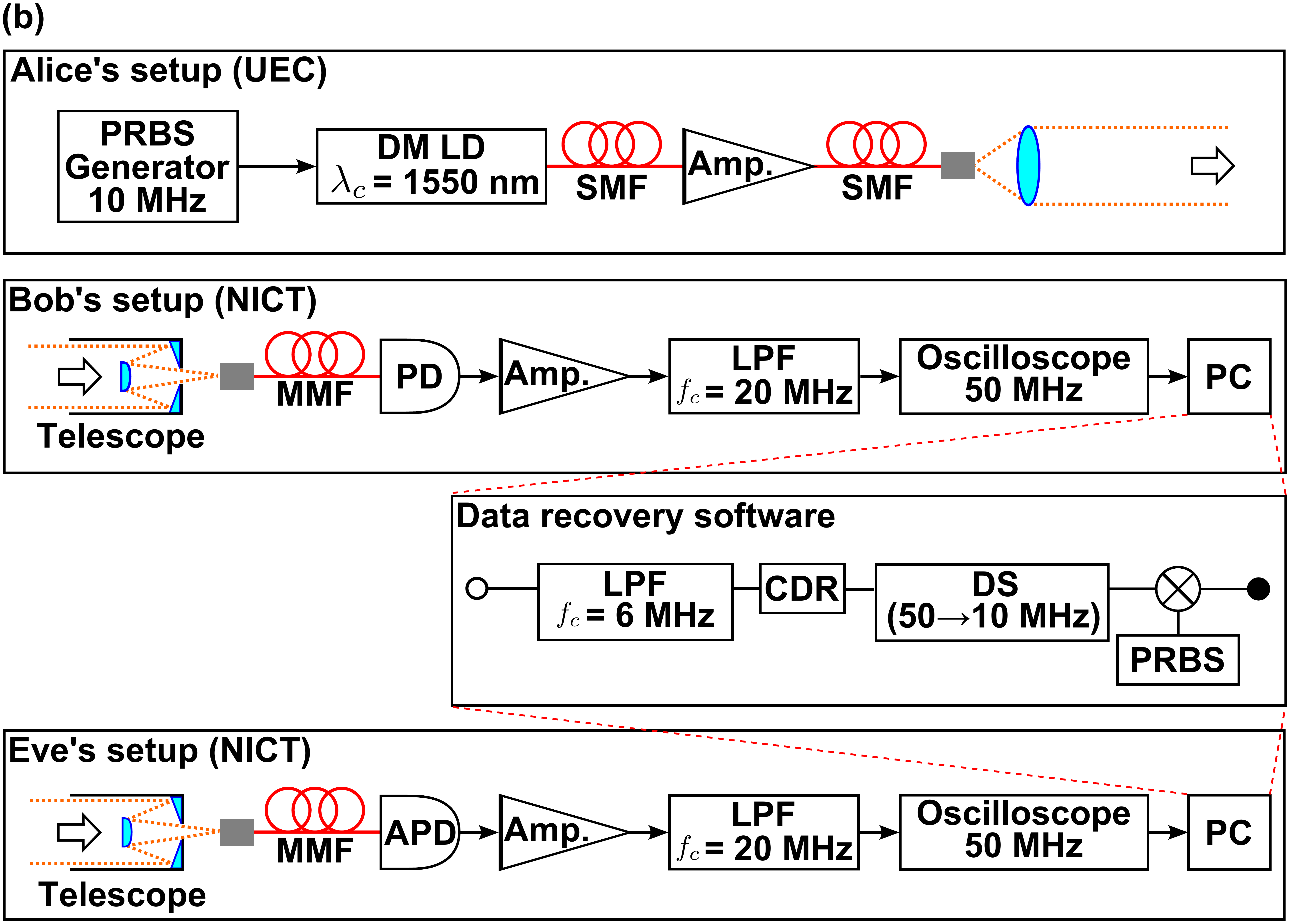}
          }
          \caption{
          (a) Overview of Tokyo FSO Testbed.  
          Alice's terminal is installed on a building roof at UEC.  
          Bob's and Eve's terminals are located on a building at NICT.  
          \copyright OpenStreetMap contributors, CC-BY-SA.  
          (b) Schematic layout of experimental setup of Alice's, Bob's, and Eve's terminals \cite{SAT}.}
          \label{expschem}
    \end{figure}
  
  A schematic layout of Tokyo FSO Testbed is shown in Fig. \ref{expschem}(a).  
  We set Alice in an all-weather telescope dome on the rooftop of a building in the University of Electro-Communications (UEC) at Chofu of Japan ($35^{\circ}39^{'}28.8^{''}$N, $139^{\circ}32^{'}39.5^{''}$E).  
  In the National Institute of Information and Communications Technology (NICT) at Koganei ($35^{\circ}42^{'}24.2^{''}$N, $139^{\circ}29^{'}19.3^{''}$E), 
  we set Bob's receiver in the sixth floor of a building.  
  All optical components are located on a high-precision motorized gimbal.  
  On the rooftop of the building which is just above the sixth floor, we set a container type terminal which takes the role of Eve.  
  This terminal consists of an all-weather scanner on the top of the container.  
  Receiver optics and electronics are located on an optical breadboard inside of the container.  
  These facilities form an FSO link with a straight-line distance of 7.8 km.  
  
  In Fig. \ref{expschem}(b), we show an overview of the optical and electrical components in our testbed.  
  The light source is a narrow linewidth direct-modulated laser diode (DM LD, Sense Light Semiconductors DL-BF10-CLS101B-S1550: band width less than 50 kHz at CW operation mode) 
  with a central wavelength $\lambda_c$ of 1550 nm.   
  This wavelength is selected since it suffers from less free-space attenuation \cite{Kimprop} and meets the eye-safety regulations \cite{Navyeyesafe}. 
  A signal is in the format of a 10 MHz pseudorandom binary sequence (PRBS) with length of $2^{15}-1$, 
  and the modulation scheme is Non-Return-to-Zero on-off keying.  
  The signal light is coupled into a fiber collimator (aperture diameter of 10 mm and divergence angle of 1.0 mrad) via a single mode fiber (SMF) and expanded into an approximately 5.5 mm beam.  
  The laser is driven by direct modulation mode, and the average output power is set to be 100 mW.  
  The collimator is mounted on a motorized gimbal driven by a high-resolution stepper motor.  
  
  At Bob's terminal, a fraction of the signal beam spot whose diameter is approximately $8$ m  
  is coupled into a Cassegrain telescope (aperture diameter of 111 mm and focal length of 800 mm) which collimates the beam down into 10 mm in diameter.  
  Then, the beam is focused into a 200 $\mathrm{\mu}$m multimode fiber (MMF) and finally sent to a photodiode detector (PD, Terahelz Technology Inc. TIA-525 optical receiver) whose noise equivalent power (NEP) is 3.0 pW/$\sqrt{\mathrm{Hz}}$.  
  The total optical loss of Bob's system, including the attenuation due to the window glass, is estimated to be -14dB.  
  At Eve's terminal, the signal beam is tapped with a Cassegrain telescope (aperture diameter of 100 mm and focal length of 2000 mm).  
  The intensity of the beam is measured by an avalanche photodiode detector (APD, Laser Components A-CUBE-I200-10) with higher sensitivity (NEP is 160 fW/$\sqrt{\mathrm{Hz}}$) than Bob's detector.  
  The total optical loss of Eve's system was measured to be -9dB.  
  The comparison of the NEPs of the detectors and the total optical losses between the terminals corroborates that Eve's receiver system is much more sensitive than Bob's one.  
  This fact allows us to emulate a reasonable situation where Eve's receiver is much better than Bob's one.  
  In addition, we can emulate more various wiretap channel conditions by directing Alice's beam to several positions between Bob and Eve.  
  
  At both terminals, the photodetector signal is amplified by a preamplifier (Hamamatsu C6438) and then sent to a USB oscilloscope (sampling rate of 50 MHz and bandwidth limit of 20 MHz by a low-pass filter (LPF)) for A/D conversion.  
  Digitized data are then sent to a computer.  
  In order to identify the transmitted signal from the received data, Bob and Eve independently perform an off-line PC-based data recovery process of which the flow chart is shown in the middle panel of Fig. \ref{expschem}(b).  
  First, high frequency noise is filtered out by a LPF (cutoff frequency $f_c$ of 6 MHz) from the input sequence.  
  Then, an accompanying clock signal is generated via a clock data recovery (CDR) process, 
  and the data are down sampled (DS) into 10 MHz of repetition rate by referencing the clock data.  
  Finally, to perform the frame synchronization, cross-correlation between the down sampled data and the original PRBS sequence is calculated.  
  In the experiment, the above-mentioned flow is implemented via LabVIEW.  

\section{Results and analysis} \label{sec4}
  \subsection{Configuration of the experiment}
      \begin{figure}[tb]
        \centering
        \includegraphics[width=12cm, bb= 0 0 1407 735]{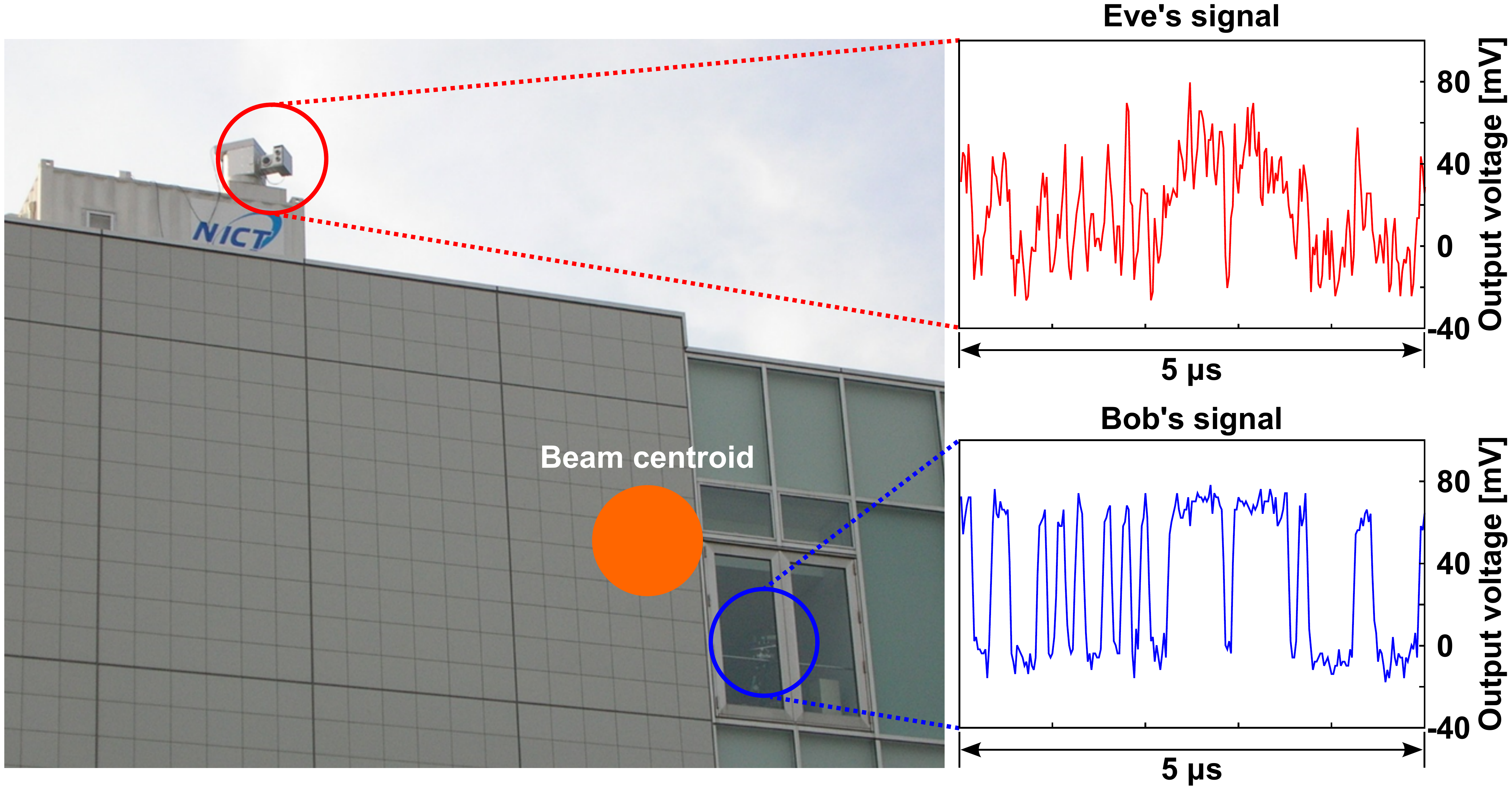}
        \caption{Experimental configuration of FSO transmission campaign held on 17 November 2015,  
                 and typical waveforms received by Eve (upper) and Bob (lower) over 5 $\mu$s. 
                 The data are taken at 14:43:00 JST. 
                 In the figure, we subtracted the DC offset of a detector from the received signal. }
                 \label{rawpower}
      \end{figure}
    
    Figure \ref{rawpower} shows our experimental configuration of FSO transmission campaign held on 17 November 2015 under cloudy Tokyo skies.  
    During the experiment, the beam centroid is put in a position closer to Bob's terminal than Eve's one as shown by red circle,   
    such that the received power at Eve would be slightly degraded compared to that at Bob.
    This geometrical configuration allows us to emulate typical wiretap channel model for satellite-to-ground laser communications where Eve attempts to tap the lobe of the beam footprint.
    Although our horizontal link cannot precisely emulate the realistic fading-induced scintillation in the vertical link of satellite-to-ground laser communications
    (the former is in one atmospheric layer while the latter is affected by several atmospheric layers with different scintillation effects),
    we can derive the basic principles of the channel estimation and the design of secure message transmission systems for a generic FSO link.
    
    We conducted the experiments in 5 time periods, namely, 14:43 - 14:46, 15:57 - 16:00, 16:33 - 16:36, 17:37 - 17:40, and 18:10 - 18:13 in JST.  
    We note that 16:33 JST was the sunset time on the day.  
    In each time period, we made 10 times of transmission and $2 \times 10^6$ bits of the PRBS are transmitted in each transmission with 200 ms duration.  
    To characterize the instantaneous secrecy rate $R_\mathrm{S,i}$, 
    we divide the duration of each 200 ms transmission into 50 of the 4 ms slot which includes $2 \times 10^5$ samples corresponding to $4 \times 10^4$ bits.
    In this time slot, the channel realizations $h_B$ and $h_E$ seem to be roughly constant hence the coherence time of the fading channels is in the order of ms.  
    Moreover, samples of statistically sufficient size are included in the duration.  
    The effect of atmospheric turbulence is nicely captured as the variation of the instantaneous secrecy rate at each time slot. 
    
    In this transmission campaign, we observed a pointing deviation of the received peak power, which can be compensated by rearranging the angles of the receiver telescopes every hour.
    We observed a typical deviation rate of 0.2 mrad/hour, which may be attributed to the refraction effect due to the varying and nonuniform air temperature 
    and to the thermal expansion of the building in which the two receivers are located.
    In the time scale of each transmission (a few minutes), the FSO link stays in the same condition.
    
  \subsection{Temporal variation of instantaneous secrecy rate}
      \begin{figure}[tb]
        \centering
        \includegraphics[width=12cm,bb=0 0 798 736]{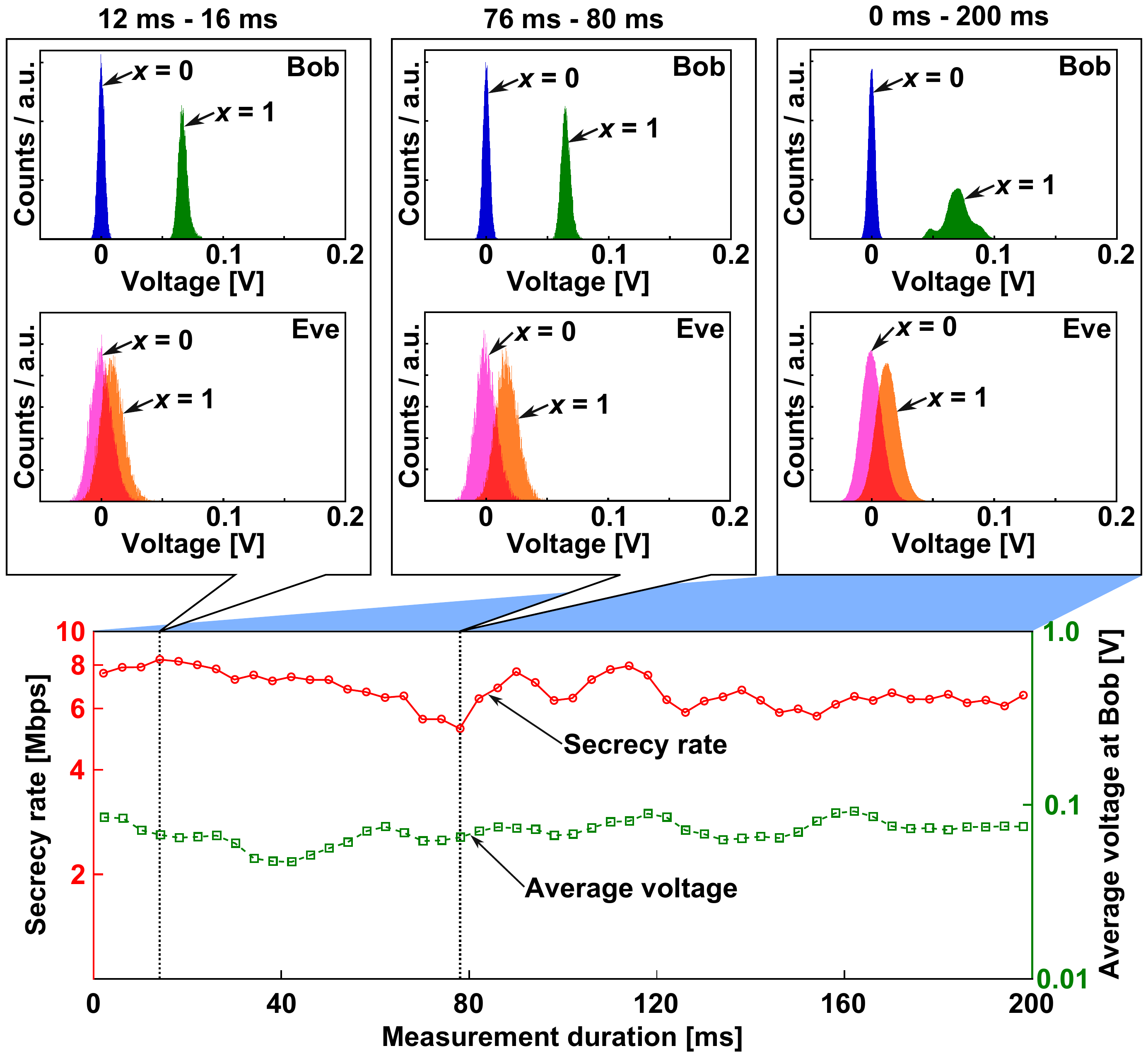}
        \caption{Temporal variation of instantaneous secrecy rate $R_\mathrm{S,i}$ (solid line) and the average output voltage (dotted line) for the experimental data at 17:37:00 JST, 
        the late evening time about an hour after the sunset, on 17 November 2015.  
        In each time slot, the measurement duration is 4 ms and $4 \times 10^4$ bits are contained.  
        Two upper left insets are the histograms of the output voltage for the best case (between 12 ms and 16 ms) and the worst case (between 76 ms and 80 ms).  
        The upper rightmost inset is the output voltage histogram for the whole period of the 200 ms transmission.  
        Width of histogram bins are 0.3 mV both for Bob's and Eve's data (see Appendix A).  
        In the histogram, we subtracted the DC offset of the detector from the received signal.  } \label{secrecyrate}
      \end{figure}
    
    In Fig. \ref{secrecyrate}, we show the temporal variation of the instantaneous secrecy rate $R_\mathrm{S,i}$ (solid line) 
    and the average output voltage (dotted line) for the typical 200 ms FSO transmission at 17:37:00 JST, 
    in the late evening time about an hour after the sunset.  
    This time period is a typical case where the fading-induced scintillation is not so heavy.  
    Actually, the highest rate (8.30 Mbits/second (Mbps), from 12 ms to 16 ms) and the lowest rate (5.25 Mbps, from 76 ms to 80 ms) are not so far 
    different (Taking into account the 10 MHz repetition rate of the input, Bob would gain 10 Mbps of information if Eve was absent.).
    Moreover, the fluctuation of the average output voltage at Bob (dotted line) seems not to have a direct correlation with the instantaneous secrecy rate $R_\mathrm{S,i}$ (solid line).  
    
    In the two left upper insets of Fig. \ref{secrecyrate}, we show the histograms of the output voltage of the detectors for the best and worst cases.  
    Clearly, the two peaks in Eve's received power histograms overlap with each other while the peaks in Bob's one are perfectly separated in both insets.    
    It turns out that under the condition of this time period, we can potentially transmit at most 5.25 Mbps of information with perfect secrecy.  
    We also show the histograms of the output voltage for the whole period of 200 ms transmission in the upper rightmost inset of Fig. \ref{secrecyrate}.  
    The spectra of these histograms are not so broader even compared to those of 4 ms time slot.  
    In general, the stronger the light fluctuation is, the broader the spectrum of the intensity distribution becomes \cite{andrewlaserprop}, 
    but Fig. \ref{secrecyrate} is not the case.  
    
      \begin{figure}[tb]
        \centering
        \includegraphics[width=12cm,bb=0 0 820 733]{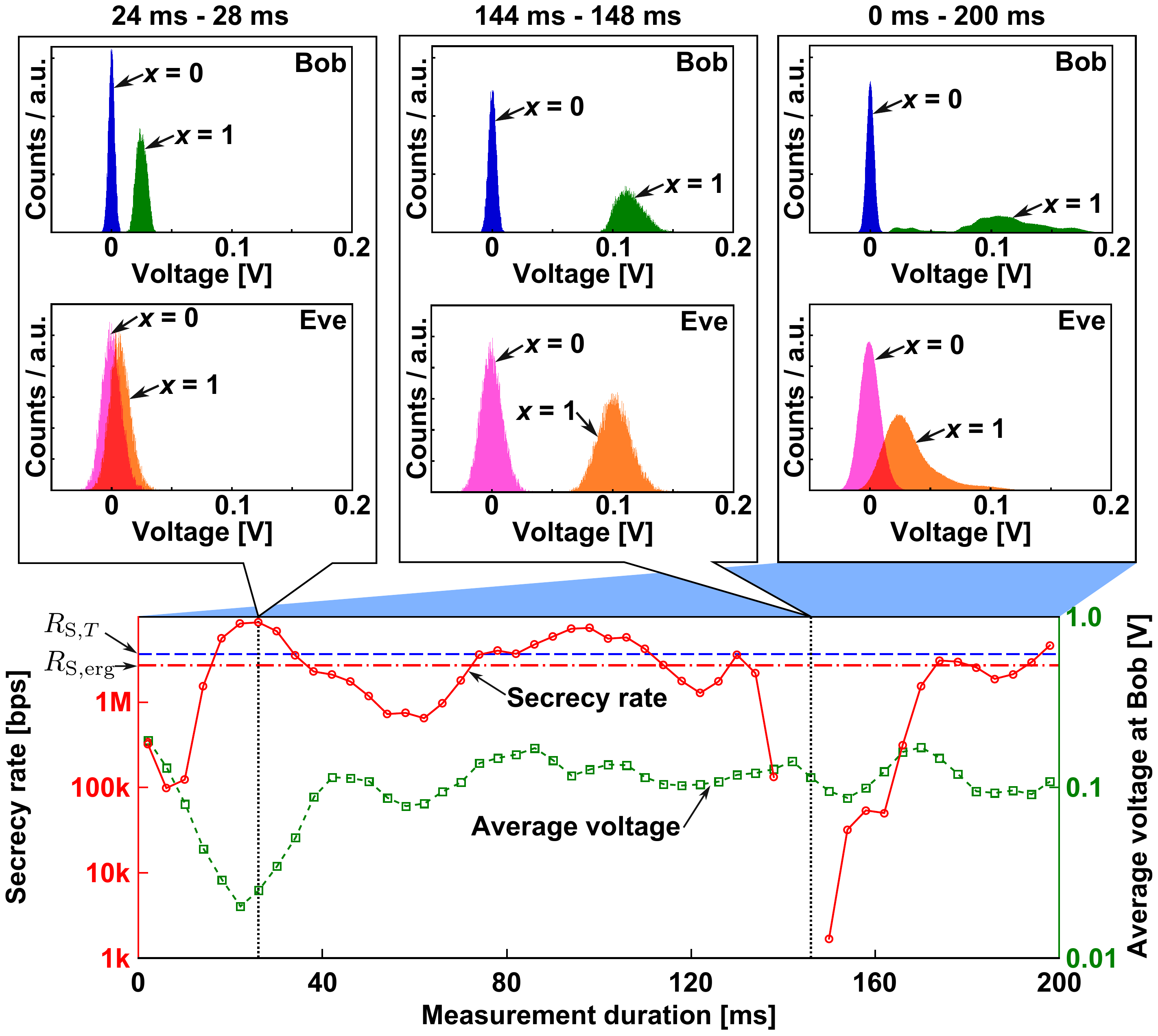}
        \caption{Temporal variation of instantaneous secrecy rate $R_\mathrm{S,i}$ (solid line) and the average output voltage (dotted line) for the experimental data at 16:34:20 JST,
        just one minute after the sunset time, on 17 November 2015.  
        In each time slot, the measurement duration is 4 ms and $4 \times 10^4$ bits are contained.  
        For comparison, the ergodic secrecy rate $R_{\mathrm{S,erg}}$ (chain line) and the long span secrecy rate $R_\mathrm{S,T}$ (dashed line) are also shown.  
        Two upper left insets are the histograms of the output voltage of the detectors for the best case (between $24$ ms and $28$) and the worst case (between $144$ ms and $148$ ms).  
        The upper rightmost inset is the output voltage histogram for the whole period of the 200 ms transmission.  
        Width of histogram bins are 0.3 mV both for Bob's and Eve's data.  
        In the histogram, we subtracted the DC offset of the detector from the received signal.  } \label{secrecy2}
      \end{figure}
    
    Figure \ref{secrecy2} shows the temporal variation of the instantaneous secrecy rate $R_\mathrm{S,i}$ (solid line) and the average output voltage (dotted line) for the typical 200 ms FSO transmission
    at 16:34:20 JST, just one minute after the sunset time.  
    Contrary to Fig. \ref{secrecyrate}, the difference between 
    the best case (8.75 Mbps, from 24 ms to 28 ms) and the worst case (0 bps, from 144 ms to 148 ms) is much more distinct.   
    For the best case, the signal via the main channel is quite distinguishable while that via the wiretapper channel is hardly distinguished as seen in the upper left inset.  
    On the other hand, around the time slot of the worst case, the instantaneous secrecy rate decreases suddenly.  
    This is because the wiretapper channel is error-free and hence Eve can establish a reliable channel from Alice.  
    It is estimated that the beam centroid should have suddenly been gotten closer to Eve's one.  
    The output voltage histograms for the whole period of the 200 ms transmission are shown in the upper-right inset of Fig. \ref{secrecy2}.  
    Compared to the one in Fig. \ref{secrecyrate}, their shapes are much broader or heavy-tailed.  
    This means that the atmospheric turbulences in this time period are much more pronounced than that in the late evening time.  
    Similar to Fig. \ref{secrecyrate}, the variations of Bob's average output voltage (dotted line) seem not to have a direct correlation with the instantaneous secrecy rate $R_\mathrm{S,i}$ (solid line).  
    Therefore, even in the present receiver configuration, based on the Eve-near-Bob scenario in \cite{LopezMartinez}, the possible spatial correlation between Eve's and Bob's channels does not show a significant impact on the secrecy rate.
    
    Figures \ref{secrecyrate} and \ref{secrecy2} indicate that Alice and Bob should find good atmospheric conditions so as to avoid a sudden fatal information leakage.  
    If there is a good correlation between Bob's and Eve's observations on some straightforward measure, such as the average output voltages, 
    then Alice and Bob could roughly infer an attainable secrecy rate (or its lower bound) by looking only at output voltage of Bob's detector.
    Unfortunately, however, no good correlation between the secrecy rate and Bob's average output voltage was seen in Figs. \ref{secrecyrate} and \ref{secrecy2}.  
    One way is to use Bob's output voltage histograms which are measured for a longer time (e.g., 200 ms like as in Figs. \ref{secrecyrate} and \ref{secrecy2}) than a fading time scale as the partial CSI.  
    If the histogram is broader, the secrecy rate varies and the fatal information leakage would occur, implying that Alice and Bob should avoid from such durations.  
    Thus Alice and Bob can use appropriate pilot signals, compare the histogram with accumulated data, and opportunistically choose good time durations like Fig. \ref{secrecyrate}.  
    
  \subsection{Code word over a longer time span} \label{43}
    Even in the larger fading case like Fig. \ref{secrecy2}, if a fast feed-forward mechanism could be employed,  
    one might be able to use an appropriate wiretap channel code, adapting changes of the channel states due to fading-induced scintillation.  
    This is, however, technically challenging.  
    
    Another possibility is to find a good code appropriately designed for the observations over a longer time span, just as shown in the upper rightmost inset in Fig. \ref{secrecy2}.  
    To distinguish these long span transition probabilities from instantaneous ones, 
    we shall call the former the long span transition probabilities, and denote them as $\mathrm{E}[P_{Y|X,H_B}]$ and $\mathrm{E}[P_{Z|X,H_E}]$ for Bob's and Eve's channels, respectively.  
    They are actually statistical mixtures of the instantaneous channel transition probabilities of Eq. (\ref{apostesoft}) and have much wider spread in distribution.  
    
    We can then calculate the secrecy rate for these long span transition probabilities as
      \begin{align}
        R_\mathrm{S,T} \equiv I(P_X, \mathrm{E}[P_{Y|X,H_B}]) - I(P_X, \mathrm{E}[P_{Z|X,H_E}]), \label{eergrate}
      \end{align}
    which we call the long span secrecy rate. 
    The value of $R_\mathrm{S,T}$ (3.71 Mbps) is shown as the dashed line in Fig. \ref{secrecy2}.  
    As seen, the long span secrecy rate $R_\mathrm{S,T}$ itself remains reasonably high, 
    even though there appear fatal decreases of the instantaneous secrecy rates in the observed time span.  
    In such fatal regions, Eve's channel remains error-free as shown in the upper middle inset. 
    The result of the long span secrecy rate implies that there exists a good code to deceive Eve even in such a situation 
    provided that the channel states remain as they are in Fig. \ref{secrecy2} for an even longer period such that the channel can be used many times with such a good code.  
    For example, one may spread a message onto a code word over the long spanned observation with sufficient randomization, and attain the secrecy even under the fading like in Fig. \ref{secrecy2}.  
    
    Interestingly, under the assumption that both the input probability distribution $P_X$ and the input power are fixed over the whole time period and  
    the main channel is almost error-free, 
    the long span secrecy rate $R_\mathrm{S,T}$ is slightly larger 
    than the ergodic secrecy rate $R_{\mathrm{S, erg}}$ (see Appendix B) which is just defined as the average of the instantaneous secrecy rates
      \begin{align}
        R_{\mathrm{S, erg}} \equiv \mathrm{E}[R_\mathrm{S,i}(h_B, h_E)], \label{ergrate}
      \end{align}
    and is often used to see an overall throughput in fading channels \cite{gopala}.  
    The value of $R_{\mathrm{S, erg}}$ (2.77 Mbps) is shown as the chain line in Fig. \ref{secrecy2} and corroborates the above point.  
    In order to achieve the ergodic secrecy rate $R_{\mathrm{S,erg}}$, one should employ a fast feed-forward mechanism 
    to adapt changes in a fading channel, which is technically challenging, as stated above.  
    Thus the transmission of a code word over a longer time span would be more attractive as fading-resistant techniques.    
    
  \subsection{Secrecy outage probability}
    In the previous subsection, we observed that the received signal statistics of the long term transmission serves as a partial CSI of the wiretap channel.  
    Alice and Bob can utilize this to determine whether they conduct secure message transmission or not.  
    However, they may dare to perform secure message transmission even with compromising the confidentiality.  
    In such a situation, a natural question arisen is then how is the trade-off relation between the throughput and the risk of information leakage.  
    In this case, the secrecy outage probability provides a quantitative metric.  
    
    The secrecy outage probability $P_\mathrm{S}(R_\mathrm{S,i} < R_\mathrm{th})$ is defined as the cumulative probability 
    that an instantaneous secrecy rate $R_\mathrm{S,i}$ is smaller than a given target rate $R_\mathrm{th}$.  
    This outage probability quantifies how often fatal information leakage occurs when the wiretap channel code is employed at constant rate $R_\mathrm{th}$.  
    Obviously, the secrecy outage probability is a monotone increase function of target rate $R_\mathrm{th}$, which indicates the trade-off relation between security and throughput.  
    
      \begin{figure}[tb]
        \centering
        \includegraphics[width=12cm, bb=0 0 850 473]{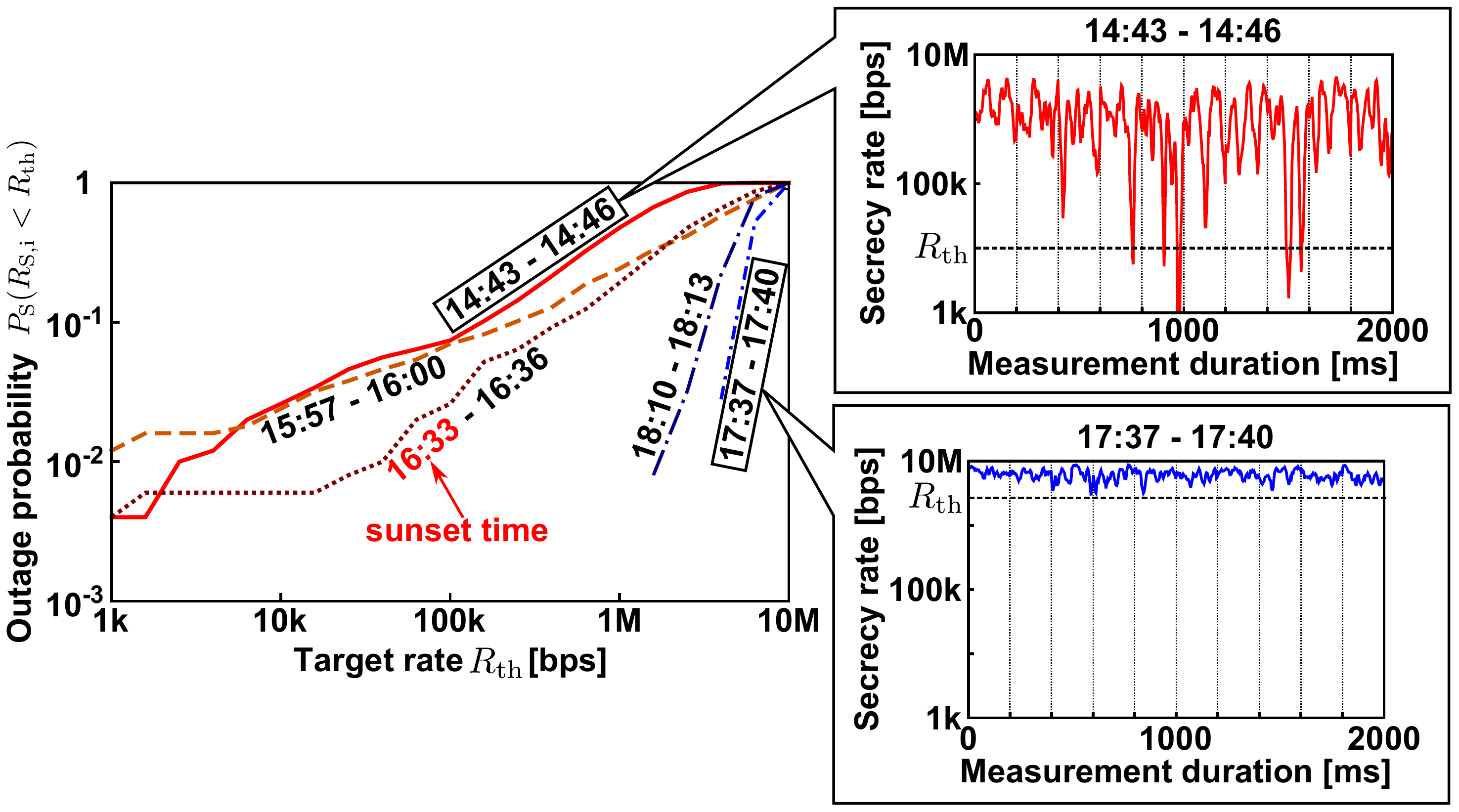}
        \caption{Secrecy outage probability $P_\mathrm{S}(R_\mathrm{S,i} < R_\mathrm{th})$ as a function of target rate $R_\mathrm{th}$ for 5 campaign periods on 17 November 2015.  
        In each time period, 10 independent 200 ms FSO transmissions (totally 20 Mbits), namely, 500 of 4 ms FSO transmission is contained.  } \label{outage}
      \end{figure}
      
      \begin{table}[tb]
        \centering 
        \caption{{\bf The mean values of the scintillation index $\sigma^2_I$ and refractive-index structure constant $C^2_n$ for each campaign period$^a$.} }
        \begin{tabular}{|c|c|c|} \hline
          Campaign period                             & Mean $\sigma^2_I$ & Mean $C^2_n$ [$\mathrm{m}^{-2/3}$]\\ \hline
          14:43 - 14:46                               & 0.076             & $2.17 \times 10^{-16}$ \\ \hline
          15:57 - 16:00                               & 0.408             & $1.16 \times 10^{-15}$ \\ \hline
          \parbox{5.5em}{16:33 - 16:36 (sunset time)} & 0.168             & $4.79 \times 10^{-16}$ \\ \hline
          17:37 - 17:40                               & 0.034             & $9.69 \times 10^{-17}$ \\ \hline
          18:10 - 18:13                               & 0.044             & $1.26 \times 10^{-16}$ \\ \hline
        \end{tabular}
        \\ {\footnotesize $^a$To calculate $\sigma^2_I$ and $C^2_n$ in this table of each 200 ms transmission, we selected the event where the light source is on.}
        \label{turbulence}
      \end{table}
    
    Figure \ref{outage} depicts the outage probability $P_\mathrm{S}(R_\mathrm{S,i} < R_\mathrm{th})$ for 5 campaign periods.  
    In each time period, 10 independent 200 ms transmissions (totally 20 Mbits) are contained.  
    The instantaneous rate $R_\mathrm{S,i}$ is calculated in the duration of 4 ms.  
    The time variations of the instantaneous secrecy rate are illustrated in right panels for 14:43 - 14:46 JST and 17:37 - 17:40 JST.  
    
    The behavior of the outage probability shown in Fig. \ref{outage} is well reflected in 
    the behaviors of the scintillation index $\sigma^2_I$ and the refractive-index structure constant $C^2_n$ \cite{andrewlaserprop} shown in Table \ref{turbulence}.
    For example, before the sunset time, $C^2_n$ indicates a larger value over $10^{-16}$.  
    On the other hand, one hour after the sunset time (17:37 - 17:40), $C^2_n$ becomes the smaller value below $10^{-16}$, 
    and $C^2_n$ turns to be a slightly larger value after further 30 minutes later (18:10 - 18:13).  
    Such a temporal suppression of fading-induced scintillation one hour after the sunset time has already been observed in the past experiment held in NICT \cite{arisa}.
    
    As shown in Fig. \ref{outage}, the secrecy outage probability is almost negligible even when the target rate is set to be more than 1 Mbps in the late evening time (17:37 - 17:40),  
    meaning that the perfect secure transmission at high throughput is possible.  
    On the other hand, before the sunset time (14:43 - 14:46), the outage probability still remains to be 0.01 even if the target rate decreases to 10 kbps.  
    Such kind of larger outage probability alerts that the perfect secrecy cannot be guaranteed solely by the wiretap channel code with $R_\mathrm{th} = 10$ kbps.  
    It advices that some backup encryption schemes in the upper layers should be activated to prepare for the worst case scenario.  
    
  \subsection{Finite length analysis}
    \begin{figure}[tb]
        \centering
          \subfloat{
            \includegraphics[width=6cm, bb =0 0 494 322]{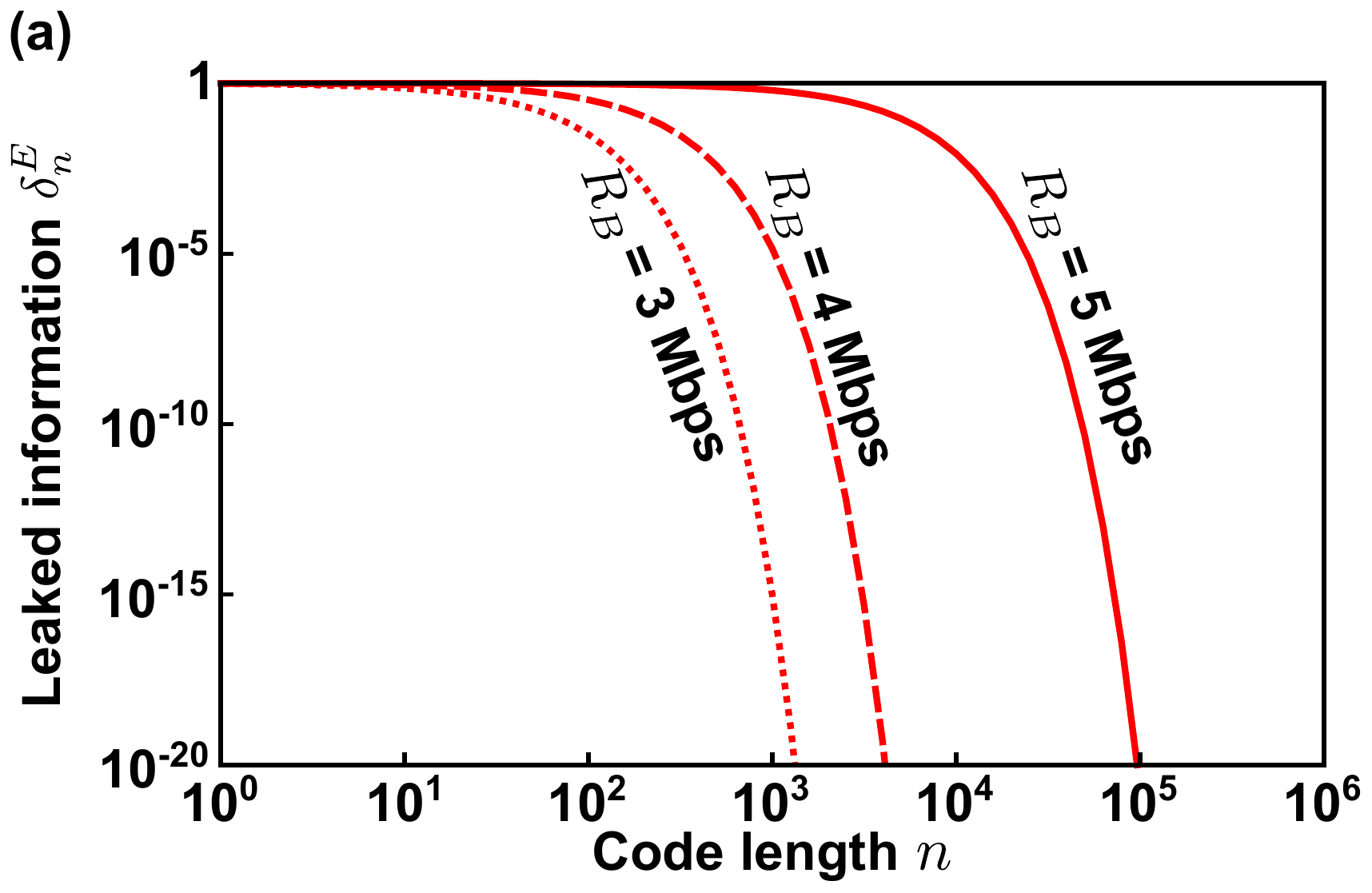}
          }
          \subfloat{
            \includegraphics[width=6cm, bb=0 0 498 343]{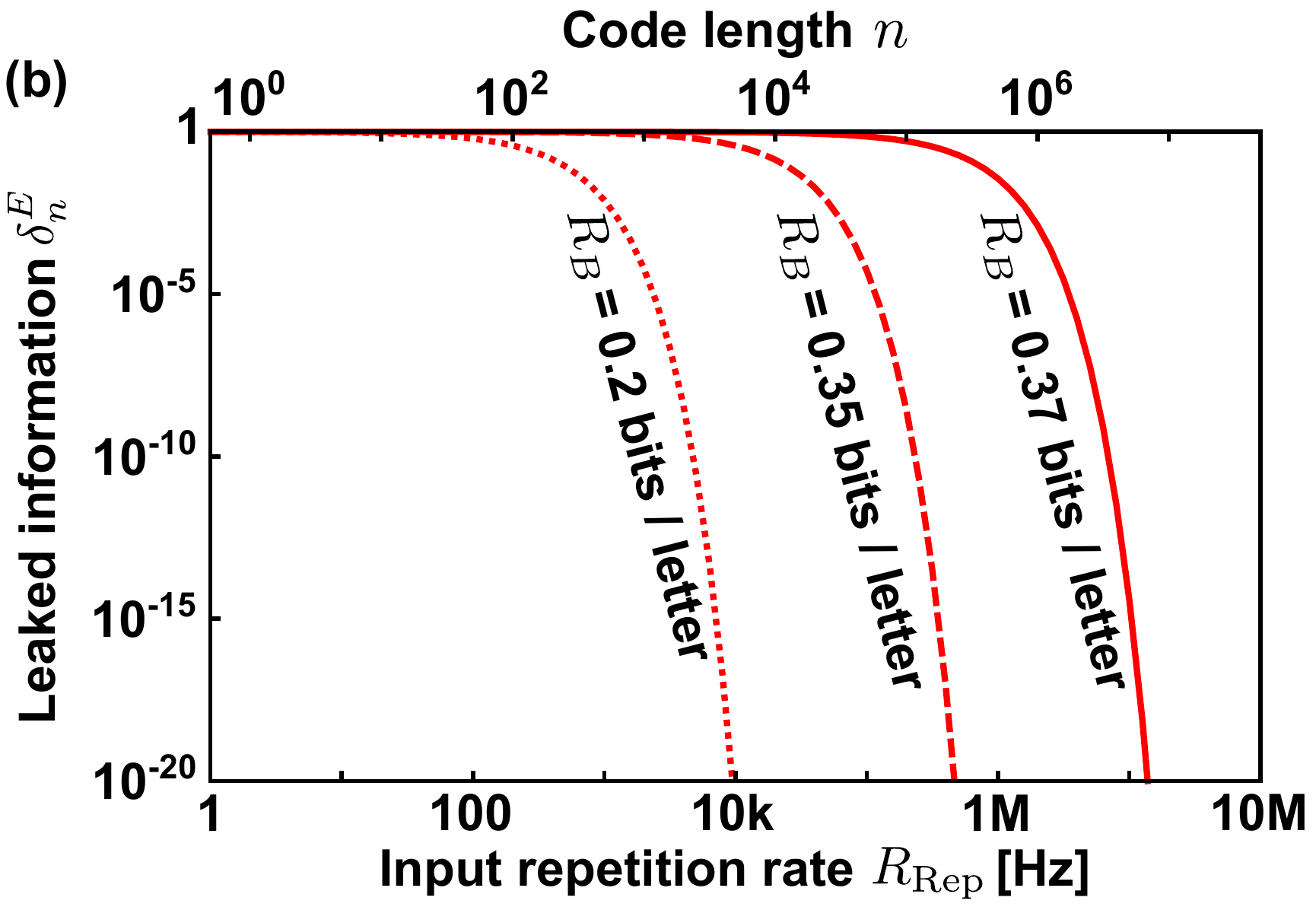}
          }
          \caption{
          (a) Code length dependence of leaked information measure $\delta^E_n$  between 76 ms and 80 ms in Fig. \ref{secrecyrate}.  
          (b) Repetition rate dependence of leaked information measure $\delta^E_n$  over the whole observation time of 200 ms in Fig. \ref{secrecy2}.  }
          \label{leakedinfo}
    \end{figure}
    
    Although the secrecy rate is regarded as a reasonable benchmark of the system, it concerns only the asymptotic limit at code length $n \to \infty$.  
    Practically, in the bounded-code-length scenario, the message rate $R_B$ cannot be arbitrarily close to the secrecy rate as well as the information leakage cannot be completely diminished.  
    Thus, in order to design a practical code with the perfect secrecy, the message rate $R_B$ should be chosen much lower than the secrecy rate 
    and the necessary code length $n$ for the required secrecy criteria should be known.  
    This motivates researchers \cite{HES, csisres, hayashi} to introduce the secrecy exponent $\Hc$ (see Appendix C),  
    which is a stronger characterization showing how fast the leaked information decreases.
    Actually, in \cite{numsec}, through the upper bound on the leaked information measure $\delta^E_n \le e^{-n \Hc}$, where the leaked information measure $\delta^E_n$ is measured by a statistical distance between distributions \cite{HES, numsec},  
    the code length dependence of $\delta^E_n$ has been investigated in the idealistic fading free model, or constant channel gain model.  
    In what follows, we consider the application of $\Hc$ on atmospheric fading channels.  

    First, we consider the case with the weaker fading case, such as in Fig. \ref{secrecyrate}.  
    In this case, the fluctuation of the instantaneous secrecy rate is also weak.  
    We may select the worst time slot, design a code for it, and apply it for the whole interval such as 200 ms in Fig. \ref{secrecyrate}. 
    Now, our purpose is to investigate what code length is required in this time slot.  
    In Fig. \ref{leakedinfo}(a), we show the code length dependence of the leaked information criteria $\delta^E_n$ on the time slot of the worst case (from 76 ms to 80 ms) in Fig. \ref{secrecyrate}.  
    Clearly, as $R_B$ decreases (or the randomness rate $R_E$ increases, since we assume that sum of the rates $R_B + R_E$ is fixed), $\delta_n^E$ decreases 
    faster, attaining a given criterion $\delta^E_n$ with a shorter code length.  
    When Alice and Bob set $R_B$ to be 3 Mbps (the dotted line in Fig. \ref{leakedinfo}(a)), $\delta_n^E < 10^{-20}$ can be obtained by a code with $n=10^3$.  
    Since this value serves as the upper bound over the whole time period, Alice and Bob reasonably achieve the secure message transmission with fixing the message rate $R_B = 3$ Mbps and the code length $n = 10^3$.  
    For the curve of $R_B = 4$ Mbps (dashed line), the required code length for $\delta_n^E < 10^{-20}$ is $n=10^4$ and still reasonable.  
    On the other hand, as we raise the rate up to $R_B = 5$ Mbps (solid line), which is close to the instantaneous secrecy rate $R_\mathrm{S,i} = 5.25$ Mbps (see Fig. \ref{secrecyrate}), 
    $n = 10^5$ of code word is required for $\delta_n^E = 10^{-20}$.  
    However, considering that the repetition rate is 10 MHz and the time slot duration is 4 ms, this code length is out of a consistent design.  
    
    Next and finally, we discuss finite length analysis in the larger fading case like in Fig. \ref{secrecy2}.  
    In Subsection \ref{43}, we have already discussed a fading resistant technique based on a code word over a longer time span, 
    i.e., the whole observation time of 200 ms. 
    The input repetition rate was set as 10 MHz.  
    This means that there are $2 \times 10^6$ symbols in the whole span, and the code length is also $2 \times 10^6$.  
    This length is however, considerably long, and readily causes large coding complexity.  
    To reduce such complexity, one must set a lower repetition rate $R_\mathrm{rep}$ for a fixed observation time $T_\mathrm{O}$, where the code length $n$ is determined as $n = R_\mathrm{rep} T_\mathrm{O}$.  
    Figure \ref{leakedinfo}(b) shows how fast the leaked information criteria $\delta_n^E$ decreases as the repetition rate $R_\mathrm{rep}$ for a given message rate $R_B = m/n$.  
    As easily imaged, if the larger message rate is required, the repetition rate must be higher for attaining a given level of secrecy, 
    and hence the code length should also be longer according to $n = R_\mathrm{rep} T_\mathrm{O}$.  
    Suppose that we set the secrecy criteria as $\delta_n^E < 10^{-20}$.  
    Then for a message rate $R_B = 0.2$ bits/letter, the repetition rate and the code length must be set roughly as $R_\mathrm{rep} = 10$ kHz and $n = 2 \times 10^3$, respectively, 
    which realizes the secure message transmission at 2 kbps.  
    For a higher message rate $R_B = 0.37$ bits/letter, 
    the repetition rate and the code length should be $R_\mathrm{rep} = 13.8$ MHz and $n = 2.77 \times 10^6$, respectively, realizing the $5.13$ Mbps secure message transmission.  
    
\section{Concluding remarks} \label{sec5}
  In this paper, we have discussed the feasibility of PHY security in real-field FSO links.  
  Using Tokyo FSO Testbed, we could gather the experimental data for various atmospheric conditions which will meet satellite-to-ground laser communications.  
  We exploited three information theoretical quantities as performance measures, the secrecy rates, the secrecy outage probability and the expected code lengths for given secrecy criteria.   
  We observed that the real conditions influence the temporal variation of the instantaneous secrecy rate; the temporal variation is stable in the late evening time, whereas it is much stronger before the sunset time.  
  When the variation of secrecy rate is stable, Alice and Bob can establish the secure message transmission with reasonable code lengths.  
  On the other hand, when the variation of secrecy rate is much heavier, Alice and Bob can assess the possibility of secure message transmission using 
  a good and longer code designed based on the long span statistics of the channel.  
  Quantitatively, they can calculate the probability of the fatal information leakage via the secrecy outage probability.  
  Combining the PHY security with upper-layer cryptographic schemes, they may be able to establish secure communication even in the heavy fading condition.  
  
  In this paper, we mainly focused on the secure message transmission via the wiretap channel.  
  As was stated, the opportunistic transmission may be impractical in the configuration of this paper since a fast adaptive optimization over input probability distribution, input power, and message rate is required.  
  However, in the delay tolerant communication such as secret key agreement \cite{AhlswedeSKA, MaurerSKA} using public channels, this opportunistic approach will work effectively.  
  In secret key agreement, Alice and Bob can opportunistically select appropriate time slots after sharing initial randomness, and apply information reconciliation and privacy amplification to the data in these time slots.  
  Moreover, the secrecy rate gives a lower bound for the achievable secret key rate.  
  The experimental result gathered via Tokyo FSO Testbed campaign tells us that the implementation of secret key agreement scheme in this testbed is straightforward.  
  Thus, the feasibility study of this scheme in real field FSO communications will be a next target of this campaign.  
  
  With a direct analysis, this study provides quantitative assessment of the key rate reduction from the eavesdropper in real channel conditions.  
  This is a strong support for the further study of this effect with a mobile attack, as for instance in the case of Eve on a drone.  
  Moreover, although our results were obtained in horizontal terrestrial propagation,  
  we believe that our feasibility study is a first step towards a realization for large-scale deployment of PHY security in FSO communication.  
  PHY security in FSO communication will open a new paradigm for the basis of high-capacity and high-altitude secure communications exploiting satellites, air planes and drones.    

\appendix
\section*{Appendix A: Bin width $\Delta$ for evaluating $I(P_X,P_{Z|X,H_E})$ with soft-decision decoding}
    \begin{figure}[!t]
      \centering
      \includegraphics[width=6cm, bb = 0 0 483 312]{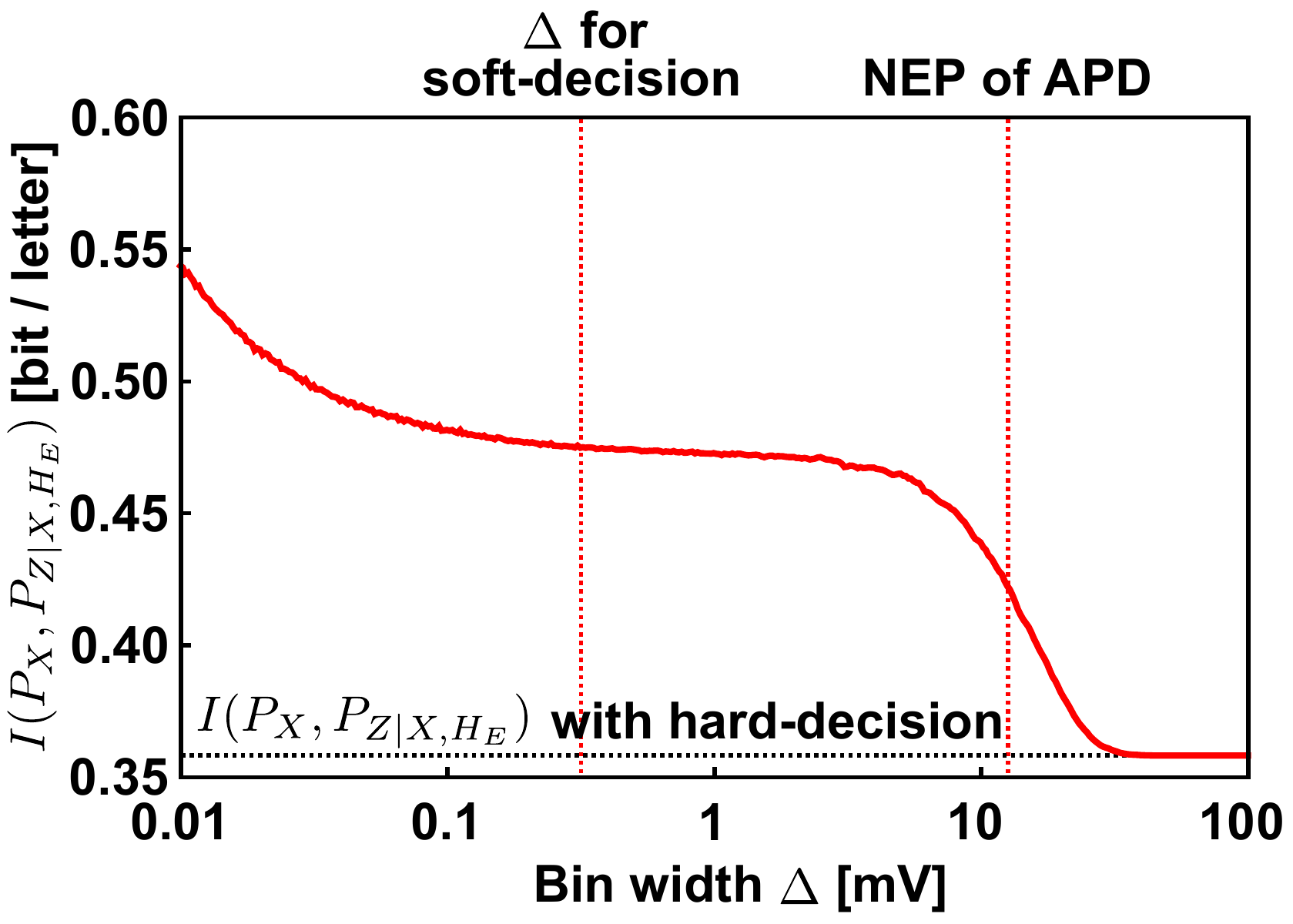}
      \caption{Bin width dependence of the mutual information $I(P_X,P_{Z|X,H_E})$ for the time slot from 76 ms to 80 ms in Fig. \ref{secrecyrate}.  }
      \label{noiseoscillo}
    \end{figure}
  
  In this appendix, we discuss the bin width $\Delta$ for evaluating the mutual information $I(P_X,P_{Z|X,H_E})$ based on soft-decision decoding.  
  The bin width for Bob's output voltage histogram (Figs. \ref{secrecyrate} and \ref{secrecy2}) was also determined by the same procedure.  
  
  Ideally, one should make the bin width $\Delta$ as finely as possible.  
  In practice, however, the finite sample size of experimental data sets the lower bound on allowed values of $\Delta$.  
  To determine such a lower bound, we investigate the bin width dependence of the mutual information $I(P_X,P_{Z|X,H_E})$ in Fig. \ref{noiseoscillo}.  
  For large $\Delta$, $I(P_X,P_{Z|X,H_E})$ stays at a constant value which corresponds to the value for hard-decision decoding.  
  On the other hand, as $\Delta$ decreases, the curve starts climbing steadily at 20 mV,
  and then reaches a plateau around at $I(P_X,P_{Z|X,H_E}) = 0.475$ bit per letter, the value for soft-decision decoding.  
  Note that the noise equivalent power (NEP) of Eve's APD, 12 mV, is located in the slope of this increase.  
  As $\Delta$ decreases further, $I(P_X,P_{Z|X,H_E})$ turns to increase again at around 0.3 mV. 
  This is due to the lack of sample size, 
  namely, one can artificially construct non-overlap distributions for the input signals 0 and 1, which is of course a fake.  
  Hence, we adopt 0.3 mV as the bin width $\Delta$ which is neither too large not to underestimate $I(P_X,P_{Z|X,H_E})$  nor too small to mitigate the effect of limited sample size.  

\section*{Appendix B: Superiority of $R_\mathrm{S,T}$ over $R_{\mathrm{S, erg}}$}
  In this appendix, we show that under the condition that both the input probability $P_X$ and input power are fixed over all coherence interval,  
  and the main channel is almost error free, 
  the long span secrecy rate $R_{\mathrm{S}, T}$ outperforms the ergodic secrecy rate $R_{\mathrm{S, erg}}$.  
  
  As is well known \cite{CKbook}, the mutual information $I(P,W)$ is a convex function of transition probability $W$ with the fixed input probability distribution $P$. 
  Thus, the following inequality holds from the Jensen's inequality \cite{CKbook};
    \begin{align}
      \mathrm{E}[I(P, W)] \ge I(P, \mathrm{E}[W]). \label{convex}
    \end{align}
  Using this inequality, we have
    \begin{align}
      R_{\mathrm{S, erg}} 
        &= \mathrm{E}[I(P_X, P_{Y|X,H_B}) - I(P_X, P_{Z|X,H_E})] \\
        & \le I(P_X, \mathrm{E}[P_{Y|X,H_B}]) - I(P_X, \mathrm{E}[P_{Z|X,H_E}]) \label{fromconvex} \\
        &= R_\mathrm{S, T},  
    \end{align}
  where Eq. (\ref{fromconvex}) follows from the assumption that Bob's channel is almost error-free and applying Eq. (\ref{convex}) on $I(P_X, P_{Z|X,H_E})$.  
  
\section*{Appendix C: Definition of secrecy exponent}
  In this appendix, we give the definition of the secrecy exponent $\Hc$.  
  The secrecy exponent $\Hc$ is defined as
    \begin{align}
      \Hc \equiv \max_{0 \le \rho < 1} \left(\phiE + \rho R_E \ln 2 \right),
    \end{align}
  where
    \begin{align}
      \phiE \equiv - \ln \sum^K_{i=1} \left(\sum_{x=\{0,1\}} P_X(x) P_{Z|X,H_E}(z^{(i)}|x,h_E)^\frac{1}{1-\rho} \right)^{1-\rho},   
    \end{align}
  with $0 \le \rho < 1$.  
  
  The secrecy exponent $\Hc$ is monotone strictly positive increasing in $R_E > I(P_X,P_{Z|X,H_E})$ and becomes $0$ for $ R_E \le I(P_X,P_{Z|X,H_E})$.  
  This is the manifestation of the tradeoff relation between information rate and secrecy \cite{HES}: 
  namely, for more secure communication, one should increase $R_E$ for the price of sacrificing the information rate $R_B$.  

\section*{Acknowledgment}
  This work was funded by ImPACT Program of Council for Science, Technology and Innovation (Cabinet Office, Government of Japan).
  The authors gratefully acknowledge inspiring discussions with Prof. Ryutaroh Matsumoto and Dr. Masahiro Takeoka.
\end{document}